\documentclass[twocolumn]{svjour3}         
\smartqed  
\usepackage{graphicx}
\usepackage[usenames]{color} 
\usepackage{amsmath,amssymb}
\begin{document}
\title{Nonlinear dynamics of a regenerative cutting process}


%
\author{Grzegorz Litak \and
        Sven Schubert \and
        G\"unter Radons 
}

%
\institute{G. Litak \at
							Department of Applied Mechanics, 
							Technical University of Lublin, 
							Nadbystrzycka 36, 
							20-618 Lublin, Poland \\
              \email{g.litak@pollub.pl}           
           \and
           S. Schubert \and G. Radons \at
              Institute of Physics, 
              Chemnitz University of Technology, 
              Reichenhainer Str. 70, 
              09126 Chemnitz, Germany \\
              \email{svs@physik.tu-chemnitz.de}}

\date{Received: date / Accepted: date}
%
%
\maketitle
\begin{abstract}
We examine the regenerative cutting process by using a single degree of freedom
non-smooth model with a friction component and a time delay term. Instead of the
standard Lyapunov exponent calculations,
we propose a statistical 0-1\,test analysis for chaos detection.
This approach reveals the nature of the cutting process
signaling regular or chaotic dynamics.
For the investigated deterministic model we are able to show a transition from chaotic to regular motion
with increasing cutting speed.
 For two values of time delay showing the different response  the results have
been confirmed
by the means of the spectral density and the multiscaled
entropy.
\end{abstract}

\section{Introduction}
A cutting process is a basic machining technology to obtain the surface of the assumed parameters.
In certain working conditions it can be disturbed by chatter appearing as unexpected waves
on 
the machined surface of a workpiece. The appearance of chatter was noticed and described by Taylor   
in the beginning of 20th century \cite{Taylor1907}. But the first approaches towards explanations  of 
this 
phenomenon 
came about 50 years later through the analysis of self-sustained vibrations  \cite{Arnold1946}, 
regenerative 
effects 
\cite{Tobias1958}, 
structural
dynamics \cite{Tlusty1963,Merit1965}, and, finally, the dry friction phenomenon 
\cite{Wu1985a,Wu1985b}. 
Consequently, elimination and stabilization of the associated oscillations have become of high interest in science 
and
technology
\cite{Altintas2000,Warminski2003,Insperger2006}. The plausible
adaptive control concept, based on relatively short time series \cite{Ganguli2007}, has been studied to gain deeper understanding.

Recently, apart from the widely developed chatter vibrations
chaotic oscillations caused 
by various system nonlinearities were predicted and detected 
\cite{Grabec1988,Tansel1992,Gradisek1998a,Gradisek1998b,Marghitu2001,Litak2002,Fofana2003,Gradisek2002,Litak2004}. 
The recent technological demand is to improve the final surface properties of the workpiece
and to minimize the production time with higher cutting speeds \cite{Stepan2003}.
Thus a better understanding of the physical phenomena associated
with a cutting process becomes necessary \cite{Martinez2009}.
In this paper we will continue the work on chaotic instabilities in cutting processes proposing the 
0-1\,test \cite{Gottwald2004,Gottwald2005,Falconer2007,Gottwald2009a,Gottwald2009b}
as a tool identifying a possible chaotic solution \cite{Litak2009}.

This paper is organized as follows. After the present introduction (Sec.\,1)
we describe the model in Sec.\,2.
In Sec.\,3 we provide the results of the simulations
and corresponding power spectral densities (PSD) 
while in Sec.\,4  the 0-1\,test is applied and, subsequently, 
the findings are confirmed by means of the multiscale entropy (Sec.\,5). 
The paper ends with conclusions (Sec.\,6).

\section{The model}
\begin{figure}[h]
\centering
\includegraphics[width=5.0cm,angle=0]{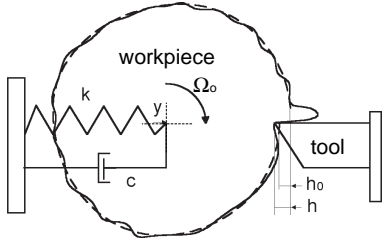}
\label{fig:model}
\caption{Physical model of a regenerative cutting process \cite{Litak2002}.}
\end{figure}
A regenerative cutting process
may exhibit a wide range of complex behavior 
due to frictional effects \cite{Warminski2003,Grabec1988}, structural nonlinearities
\cite{Pratt2001}
and delay dynamics 
\cite{Litak2002,Fofana2003,Stepan2001,Wang2006,Litak2008}.
Moreover it may also involve loss of contact between the tool and the workpiece.
The following equations model the regenerative cutting process and the mentioned properties.

After the first pass of the tool, the cutting depth can be expressed as
\begin{eqnarray}
\label{eq1}
h(t)=h_0-y(t)+y(t\!-\!\tau),
\end{eqnarray}

\noindent where $y(t\!-\!\tau)$ corresponds to the position
of the workpiece during the previous pass, and $\tau$ is the time delay scaled
by the period of revolution of the workpiece $2\pi/\Omega_0$ (Fig.\,1). 
The motion of the workpiece can be determined from the model proposed by
St\'ep\'an \cite{Stepan2001}
\begin{eqnarray}
\label{eq2}
 &&       \ddot y + 2\gamma \dot y + \omega_0^2 y =
        \frac{1}{m}{\rm sgn} (v_0\!-\! \dot y)\big(F_y(h) - F_y(h_0)\big), \nonumber \\
&&
  F_y(h)= \Theta(h) c_1 w
h^{3/4},  \\
&&
 \dot y (t^+)= -\beta \dot y (t^-), \nonumber
\end{eqnarray}
\noindent where $\omega_0=\sqrt{k/m}$ is the frequency of
free vibration, $v_0$ is the feed velocity, and $2\gamma=c/m$ is the damping coefficient.
$F_y(h)$ is the thrust force, which is the horizontal component of the
cutting force, and $m$ is the effective mass of the workpiece.
The thrust force $F_y$ is based on dry friction
between the tool and the chip. It is assumed to have
a power law dependence on the actual cutting
depth $h$ and to be proportional to the chip width $w$ and a friction coefficient $c_1$.
$\Theta(\cdot)$ denotes the Heaviside step function. The restitution parameter 
$\beta=0.75$ is associated with the impact
after contact loss, while $t^-$ and $t^+$ denotes the time instants before and after the impact.
Substituting \mbox{Eq.\,(\ref{eq1})} into  \mbox{Eqs.\,(\ref{eq2})}
we derive
a delay differential equation (DDE) for the workpiece motion $y(t)$.
Plugging its solution into \mbox{Eq.\,(\ref{eq1})}
results in the history of cutting depth $h(t)$.

\section{Simulation results}
The non-smooth model equations are solved by a simple Euler integration scheme.
The used parameters \cite{Litak2002,Litak2008} are presented in Table\,\ref{tab:1}.
\begin{table}[h!]
\caption{Parameters used in the model}
\label{tab:1}       
\begin{tabular}{ll}
\hline\noalign{\smallskip}
Parameter\hfill & Value \\
\noalign{\smallskip}\hline\noalign{\smallskip}
initial cutting depth $h_0$ & $10^{-3}$ m \\
frequency of free vibration $\omega_0$ & 816 rad/s \\
damping coefficient $c$ & $86$ Ns/m \\
effective mass of the workpiece $m$ & 17.2 kg \\
friction coefficient $c_1$ & $1.25\times10^9$ N/m$^2$ \\
chip width $w$ & $3.0 \times 10^{-3}$ m \\
\noalign{\smallskip}\hline
\end{tabular}
\end{table}
Furthermore, the feed velocity $v_0$ has been assumed to be
fairly large so that $v_0\!>\! \dot y$.
  Note that, in this case, the system nonlinearities are limited to the 
 exponential dependence of the cutting force on the chip 
thickness and to the contact loss between the tool and workpiece.

\begin{figure}[b]
  \parbox[b]{1mm}{a)}\includegraphics[width=0.45\textwidth]{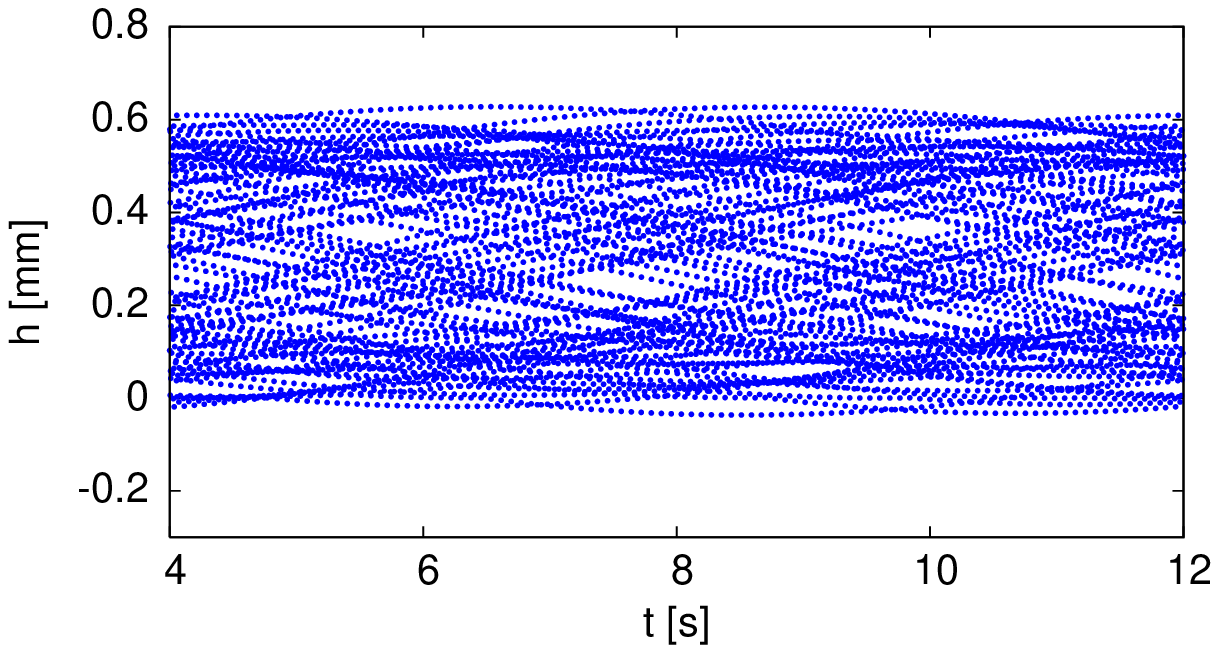} \\
  \parbox[b]{1mm}{b)}\includegraphics[width=0.45\textwidth]{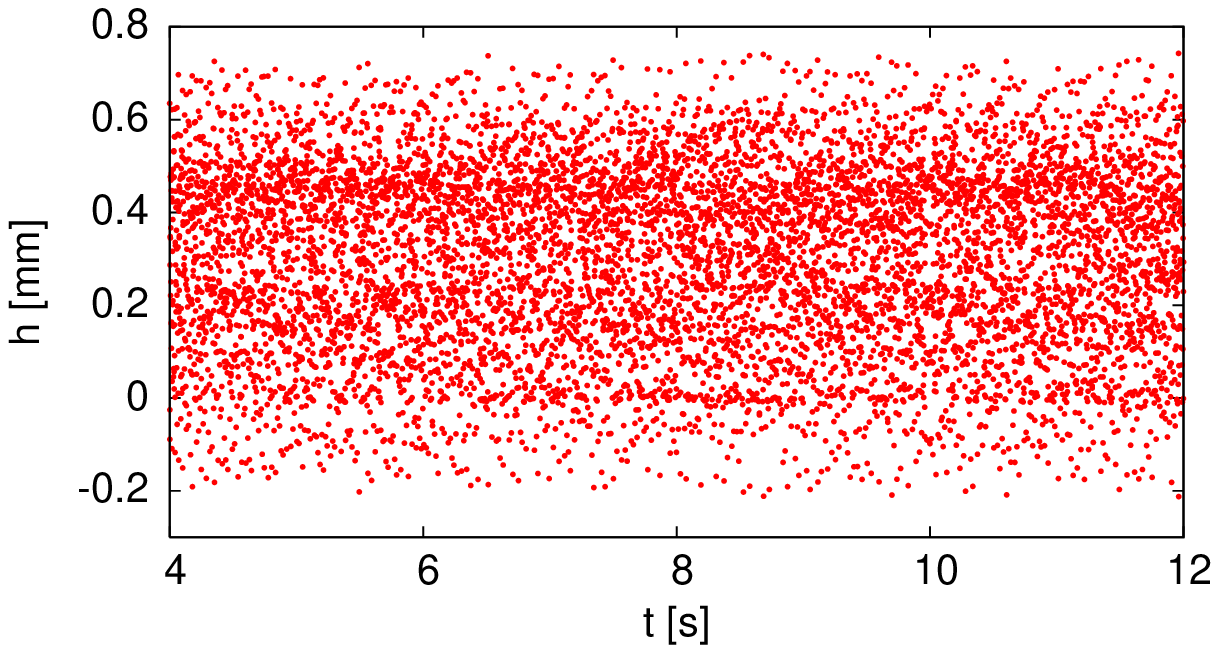}
  \caption{Time series of cutting depth $h(t)$ for \textbf{(a)} time delay $\tau = 1.8$\,ms
                and \textbf{(b)} $\tau = 2.1$\,ms (sampling time $\Delta t = 1$\,ms 
 with an integration step $\Delta t$/$10^3$=1$\mu$s)
                indicating regular and chaotic motion, respectively. The time series are plotted by
sampling points.}
  \label{fig:data}
\end{figure}

The corresponding time series for two choices of the time delay parameter $\tau = 1.8$ and 2.1\,ms 
are presented in 
Fig.\,\ref{fig:data}. These series have been plotted with points. On the first sight one can notice 
that both 
solutions are 
complex but 
the 
Fig.\,\ref{fig:data}a shows 
points grouped in selected lines while the distribution of time history points of Fig.\,\ref{fig:data}b looks 
more random. In Fig.\,\ref{fig:data}b $h$ reaches negative values that signal that the contact between 
the tool and the workpiece is lost.
%
\begin{figure}
  \includegraphics[width=0.45\textwidth]{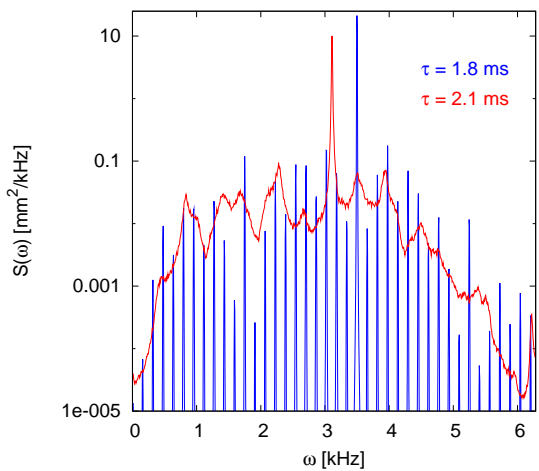}
  \caption{Power spectral density $S(\omega)$ for two chosen
                delay times. A broad band spectral density indicates chaotic /
                stochastic dynamics whereas sharp peaks imply regular motion.}
  \label{fig:PSD}
\end{figure}
The power spectral densities (PSD) of cutting depth
$S(\omega)={2\pi}\!/{T}\,|{\cal F}\{h(t)\}|^2$
for the two chosen delay times ($\tau = 1.8$\,ms and $\tau = 2.1$\,ms)%
\footnote{${\cal F}\{\cdot\}$ denotes the Fourier transform.} 
indicate a transition from regular to chaotic motion.
The sharp peaks in Fig.\,\ref{fig:PSD} belong to a high-periodic orbit 
(regular motion) whereas the broad spectrum indicates chaotic dynamics.

Both power spectra are dominated by a main peak.
In case of regular motion its position belongs to the delay time $\tau=1.8$\,ms
while in case of chaotic dynamics the time scale belonging to the peak ($t_{p}\approx 2.0$\,ms)
is smaller than the delay time $\tau=2.1$\,ms. This smaller value could be a consequence 
of a tool-workpiece contact loss.
Based on that we take a closer look on other measures to characterize the model's dynamics and
 use a 0-1\,test for chaos to display a possible transition from regular to chaotic motion with increasing 
delay time $\tau$.

\section{Application of 0-1\,test}

\begin{figure}[b]
\centerline{
\includegraphics[width=0.375\textwidth,angle=-90]{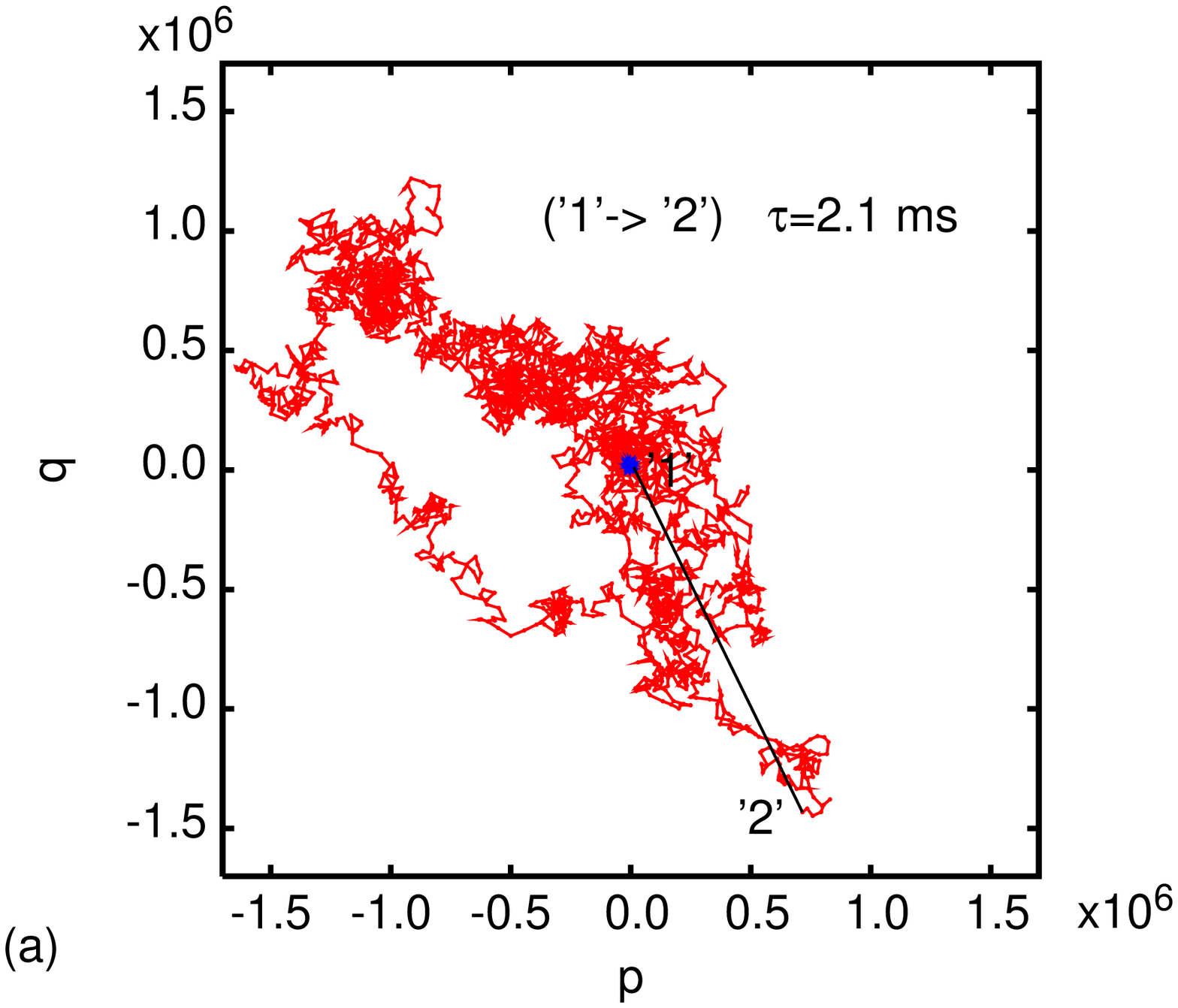}} %

\centerline{\includegraphics[width=0.375\textwidth,angle=-90]{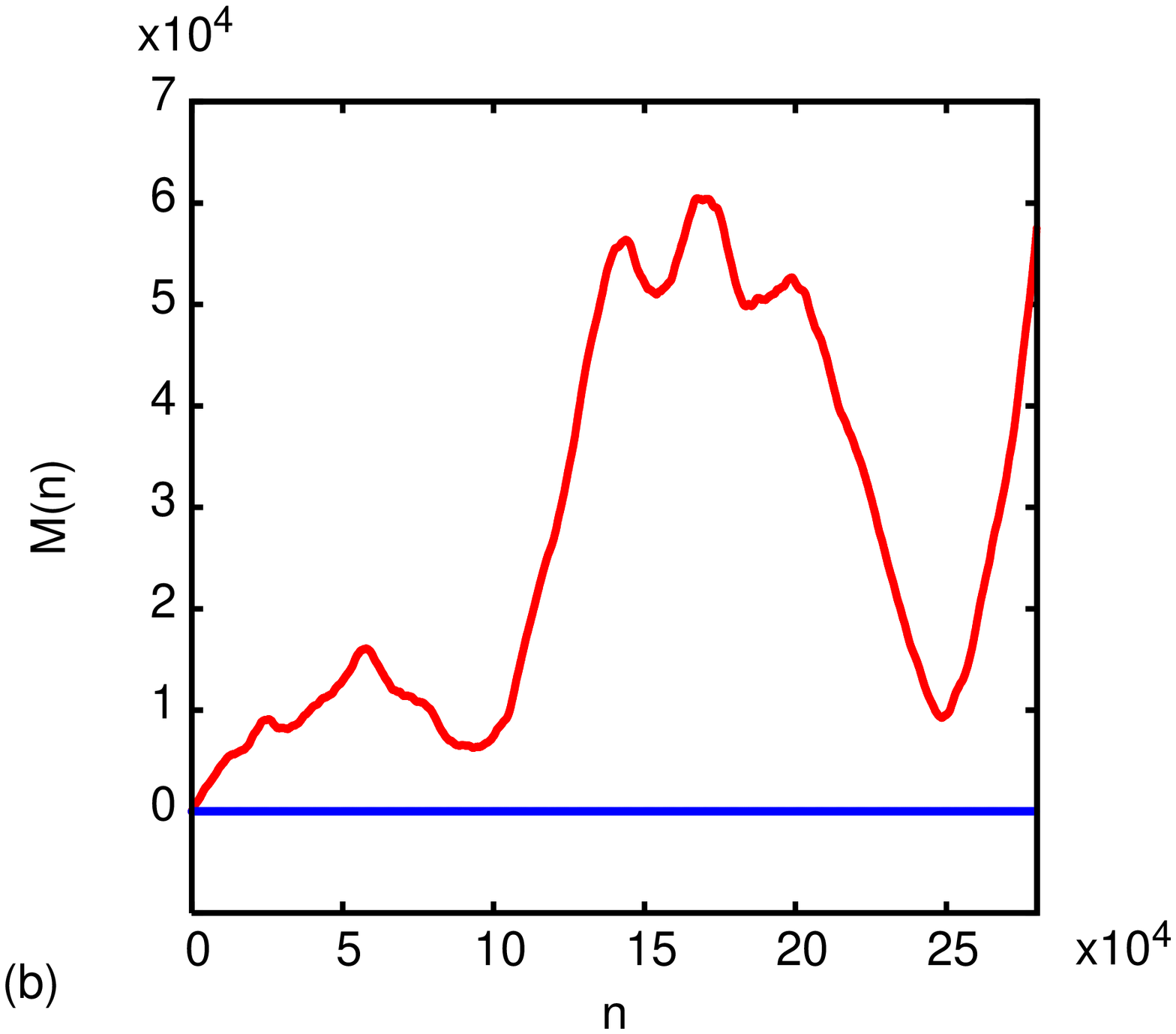}}

\caption{\textbf{(a)} For regular motion ($\tau =$ 1.8 ms) trajectories in $(p,q)$
stay bounded around the initial point '1'.
		For chaotic time series ($\tau = 2.1$ ms) trajectories in $p$ and $q$ coordinates
        show Brownian motion-like behavior. \textbf{(b)} 
        Thus the mean square displacement $M(n)$ increases with time.
		For estimation of
$p_n$ and  $q_n$, Eqs.\,(\ref{eq5}), we used $c_0=0.7$ and for
$M(n)$ and $K$, Eqs.\,(\ref{eq6}) and (\ref{eq7}), the upper limits of $N$, $n$ are
$N_{max}=40000$,
$n_{max}=280000$, respectively.
\label{fig:2}}
\end{figure}

\addtocounter{figure}{-1}
\begin{figure}
\centering
\centerline{
 \includegraphics[width=0.415\textwidth]{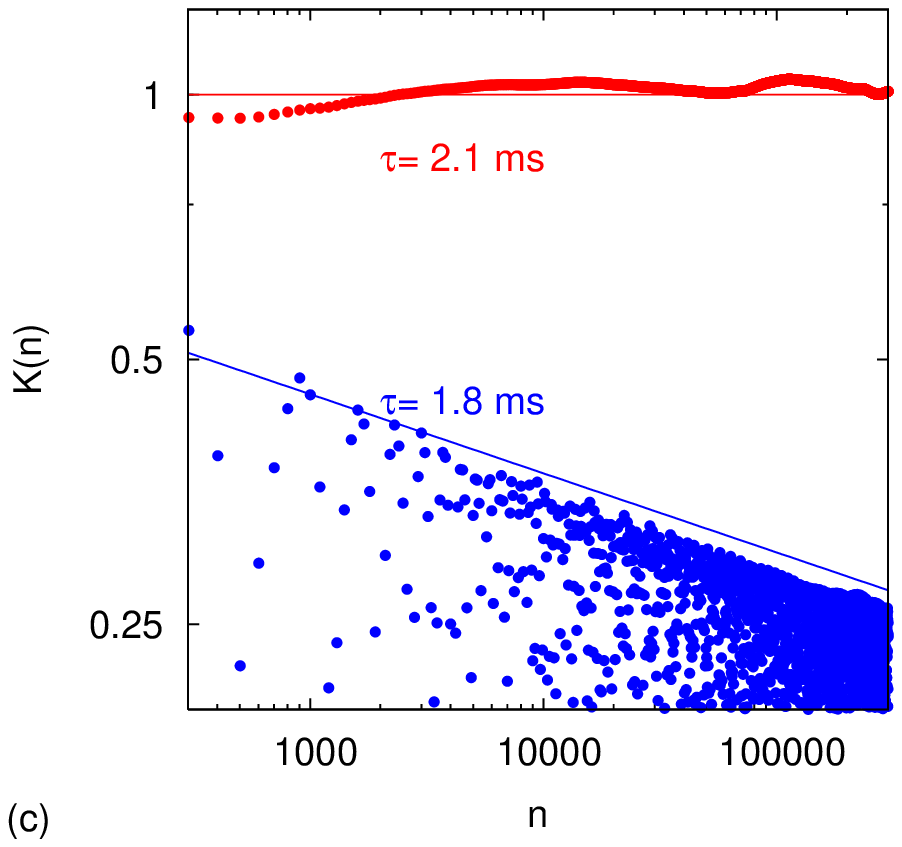}}

\caption{\textbf{(c)} For regular dynamics ($\tau = 1.8$ ms) 
	the $K$-value converges towards zero with increasing time.
	For chaotic dynamics ($\tau = 2.1$ ms) $K$ stays close to 1.}

\end{figure}

Based on the time series $\{\tilde h_j\}$
which is a discretization of the solution $h(t)$ of the DDE
normalized by its standard deviation,
we define dimensionless displacements in the $(p,q)$-plane in the following way 
\cite{Gottwald2004,Gottwald2005,Litak2009}
\begin{equation}
\label{eq5}
p_n = \sum_{j=0}^{n} \tilde h_j \cos (j c_0), \hspace{0.6cm}
q_n = \sum_{j=0}^{n} \tilde h_j \sin (j c_0),
\end{equation}
where $c_0$ is a constant.
In this way 
regular dynamics is related to a bounded motion
while any chaotic dynamics leads to an unbounded motion in the $(p,q)$-plane
\cite{Gottwald2004}, see Fig.\,\ref{fig:2}a.

To obtain a quantitative description of the examined system we perform
calculations of the asymptotic properties
defined by the total mean square displacement (MSD) $M(n)$, \mbox{Fig.\,\ref{fig:2}b},
and finally we obtain the 
growth rate $K$
in the limit of large times
\begin{eqnarray}
\label{eq6}
&& M(n)= \lim_{N \rightarrow \infty} \frac{1}{N} \sum_{j=1}^N
\big[\left(p_{j+ n}\! -\!p_j\right)^2 + \left(q_{j+ n}\! -\!q_j\right)^2\big]
, \\
\label{eq7}
&& K= \lim_{n \rightarrow \infty} \frac{\ln (M(n)+1)}{\ln n}.
\end{eqnarray}
For almost all values of the constant $c_0$ the parameter $K$ is approaching 
asymptotically 0 or 1 for regular or chaotic
motion, respectively.

Note, practically, one has to truncate the sums in \mbox{Eq.\,(\ref{eq6})}.
Thus we derived $K\approx 0.21$ 
for $\tau = 1.8$ ms and $K\approx 1.09$ for $\tau = 2.1$ ms,
which supports the first impression gained from the time series themselves, 
Fig.\,\ref{fig:data}a and b.
Note further that for delay time $\tau = 1.8$\,ms, $K$ decays 
with increasing $n$ on much smaller values, Fig.\,\ref{fig:2}c, 
which corroborates the result pointing towards regular motion.

\begin{figure}
  \centering
  \includegraphics[width=0.45\textwidth]{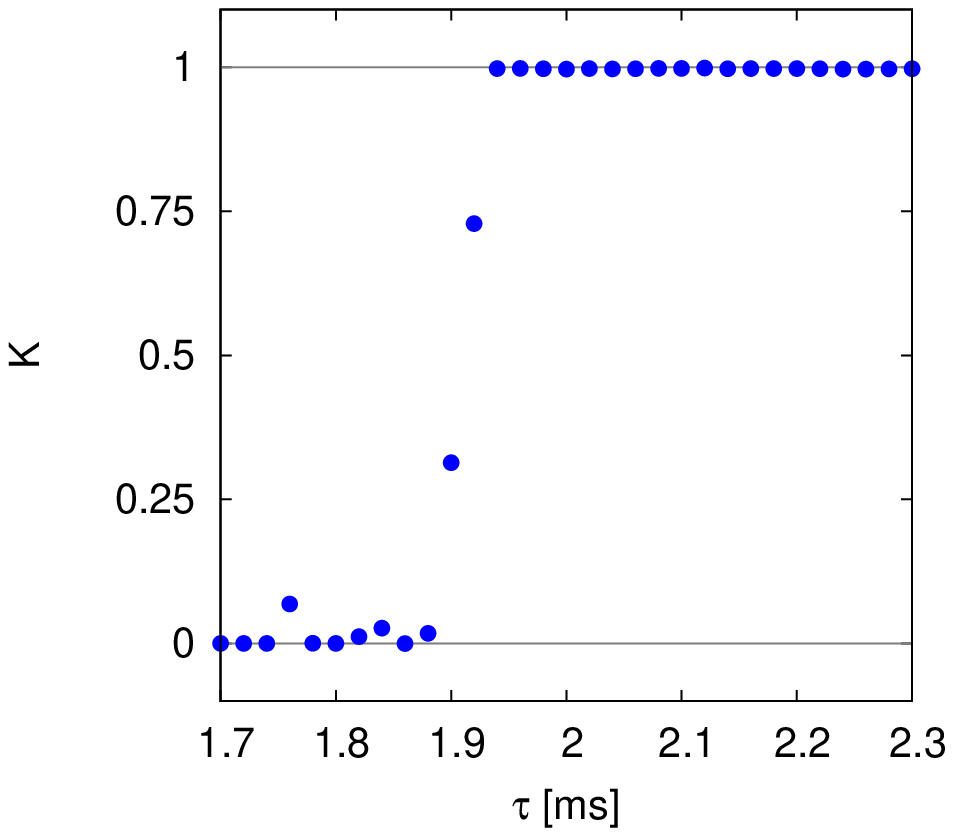} 
  \caption{$K$-values for different delay times $\tau$ indicate a transition from regular to
        chaotic dynamics in the region of 1.9\,ms.
  }
  \label{fig:Kdelay}
\end{figure}

 Note that, the parameter $c_0$ acts like a frequency in a spectral calculation, cp. Eqs.\,(\ref{eq5}). If it 
is badly chosen,
$c_0/{\Delta t}$ resonates with one frequency of the process dynamics $\tilde h(t)$.
Such a frequency belongs to a a peak in the PSD, Fig.\,\ref{fig:PSD}.
In the 0-1\,test regular motion would yield a ballistic behavior in the $(p,q)$-plane
and the corresponding quadratic growth of MSD results in
an asymptotic growth rate.
The disadvantage of the test, its strong dependence on the chosen parameter $c_0$, could be overcome by a proposed
modification. Gottwald and Melbourne  \cite{Gottwald2005,Gottwald2009a} suggest to take several
randomly chosen values of $c_0$
and compute the median of the belonging $K$-values.  Particularly, in Ref. \cite{Gottwald2009a} the
problems of averaging over $c_0$ as well as sampling the data points are
discussed extensively.
We followed this approach \cite{Gottwald2009a,Krese2012},
which improves the convergence of the test (Fig. 4c) without the consideration of longer time series,
to find the time delay $\tau$ leading to chaos   
(see Fig. \ref{fig:Kdelay}).
We defined a modified square displacement $D(n)$ which exhibits the same asymptotic growth
\begin{equation}
D(n,c_0)=M(n,c_0)-V_{osc}(n,c_0),
\end{equation}
where the oscillatory term $V_{osc}(n,c_0)$ can be expressed by
\begin{equation}
V_{osc}(n,c_0) =E[\tilde{h}]^2 \frac{1- \cos(nc_0)}{1-\cos(c_0)}, 
\end{equation}
and $E[\tilde{h}]$ denotes the average of examined time series  $\tilde{h}_i$
\begin{equation}
E[\tilde{h}]=  \frac{1}{N_{max}} \sum_{i=1}^{N_{max}} \tilde{h}_i,  
\end{equation}
where $N_{max}$ is the number of $\tilde{h}_i$ elements.
Consequently, the oscillatory behavior is subtracted from the MSD $M(n,c_0)$ and the regression analysis
 of the linear growth of $D(n,c_0)$ (Eq. 6) with increasing $n$ is performed using the linear correlation 
coefficient which determines the value of $K_{c_0}$.
\begin{equation}
K_{c_0}= \frac{{\rm cov}({\bf X},{\bf D}(c_0))}{\sqrt{{\rm var}({\bf X}) {\rm var} ({\bf D}(c_0)})},  
\end{equation}
 where  vectors ${\bf X}$=[1, 2, ..., $n_{max}$], and ${\bf D}(c_0)$= [$D(1,c_0)$, $D(2,c_0)$, ....,  
$D(n_{max},c_0)$].

 The covariance ${\rm cov}({\bf x}, {\bf y})$ and variance ${\rm var}({\bf x})$,  for arbitrary vectors ${\bf 
x}$ and ${\bf y}$ of  $n_{max}$ elements,  
and the corresponding averages  $E[x]$ and $E[y]$ respectively,
are 
defined 
\begin{eqnarray}
{\rm cov}({\bf x},{\bf y}) &=& \frac{1}{n_{max}} \sum_{n=1}^{n_{max}} (x(n)-E[x])(y(n)-E[y]), \nonumber \\
{\rm var}({\bf x}) &=& {\rm cov}({\bf x}, {\bf x}).
\end{eqnarray}

Finally, the
median is taken of $K_{c_0}$-values (Eq. 9) corresponding to 100 different values of $c_0 \in (0,\pi)$.
The results of $K$ for different delay times
$\tau$, Fig.\,\ref{fig:Kdelay},
in the window between 1.75\,ms and 2.3\,ms indicates a
transition from regular to chaotic dynamics with increasing
delay time in the region of 1.9 ms.

As a consequence we conclude that in
the investigated window increasing cutting speed leads
to a transition from chaotic chatter dynamics to regular
motion with improved surface quality.

\section{Multiscale entropy}\label{sec:MSE}
To characterize the solutions of the DDE, Eqs.\,(\ref{eq1}) and (\ref{eq2}),  
with regard to information production rate and complexity, 
we aim to calculate multiscale entropy (MSE) \cite{Costa2002}.
This method was successfully applied to analyze the complexity of biological signals \cite{Costa2002,Costa2005}.
It is suitable for short and noisy time series. 
As a consequence the chosen procedure would be applicable to experimental data as well.
We use an algorithm provided by PhysioNet \cite{PhysioNet}.
First we compute coarse-grained time series $\{x^{(N)}\}$
using non-overlapping intervals containing $N$ equidistant data points $h_i$,
\begin{equation}\label{eq:coarsegraining}
	x_j^{(N)}=\frac{1}{N}\sum\limits_{i=(j-1)N+1}^{jN} h_i . 
\end{equation}
In the next step we calculate sample entropy $S_E^{(N)}$ \cite{Richman2000} for these coarse-grained time series.
Sample entropy is the negative of the logarithm of the conditional probability
that sequences of $m$ consecutive data points 
$\textbf{x}^{(N)}_i\!=\!(x^{(N)}_i,\ldots,x^{(N)}_{i+m-1})$ and $\textbf{x}^{(N)}_j$ close to each other 
will also be close to each other when one more point is added to them.
Hence it is estimated as follows
\begin{equation}\label{eq:MSE}
	S^{(N)}_E(m,r) = -\ln\frac{U^{(N)}_{m+1}(r)}{U^{(N)}_{m}(r)},
\end{equation}
where $U^{(N)}_{m}(r)$ represents the relative frequency that a vector $\textbf{x}^{(N)}_i$
is close to a vector $\textbf{x}^{(N)}_j$ ($i\neq j$).
Close to each other in the sense that their infinity norm distance 
is less than $\varepsilon=r\sigma$. 
By $\sigma$ we denote the standard deviation of the data.
In the limit of $m\!\rightarrow\!\infty$ and $r\!\rightarrow\! 0$ sample entropy is equivalent to
order-2 R\'enyi entropy $K_2$
and is suitable to characterize the system's dynamics \cite{Grassberger1983}.
For independent variables $\{\xi\}$ the entropy follows from 
$S^{(N)}_E(m,r)=-\ln P\big(|\xi^{(N)}_i\!-\!\xi^{(N)}_j|<\varepsilon\big)$
and is independent of word length $m$.
For Gaussian white noise (GWN) 
the coarse-grained time series is known to be Gaussian distributed too. 
For small $\varepsilon$ this yields 
$S^{(N)}_E(m,r) \approx -\ln [\varepsilon/(\sqrt{\pi}\sigma^{(N)})]$.
Using that the standard deviation of the coarse-grained time series $\sigma^{(N)}$ decreases with $1/\sqrt{N}$
leads to following expression
\begin{equation}\label{eq:MSE_GWN}
	S^{(N)}_E(m,r) \approx -\ln\! \Big(\!r\sqrt{\!\frac{N}{\pi}}\Big)\;(r\rightarrow 0).
\end{equation}

\begin{figure}
  \includegraphics[width=0.5\textwidth]{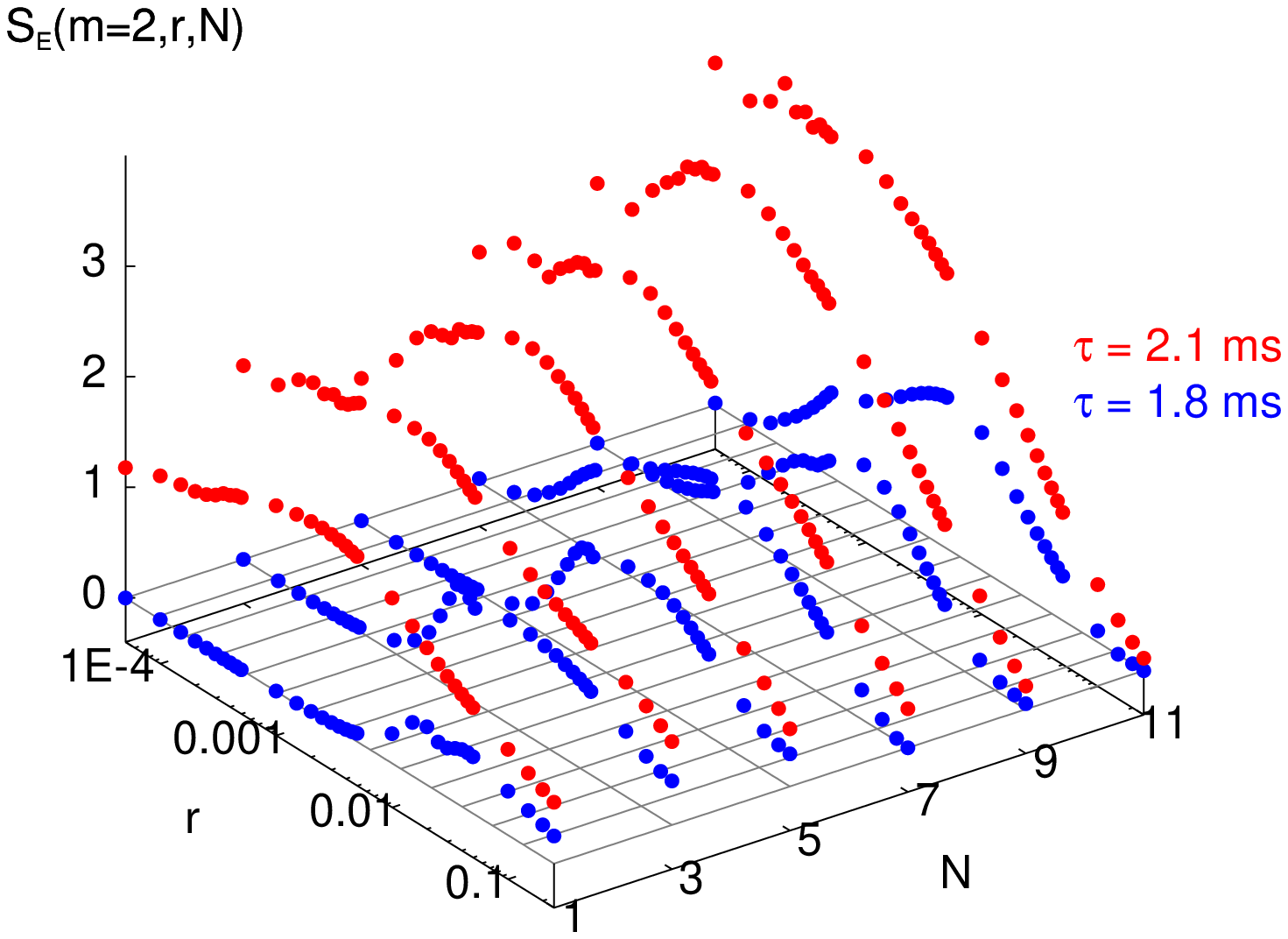}
  \caption{Multiscale entropy $S_E$ depending on scale factor
		$N$ and box size $r$. For regular motion we expect the entropy to
		approach zero with decreasing $r$. 
		For chaotic dynamics the entropy should stay finite. Since $S_E$ is
		not decreasing with scale factor significantly, it seems there is no
		characteristic time scale present in the data.}
  \label{fig:MSE2d}
\end{figure}

\begin{figure}
  \parbox[b]{1mm}{(a)}\includegraphics[width=0.43\textwidth]{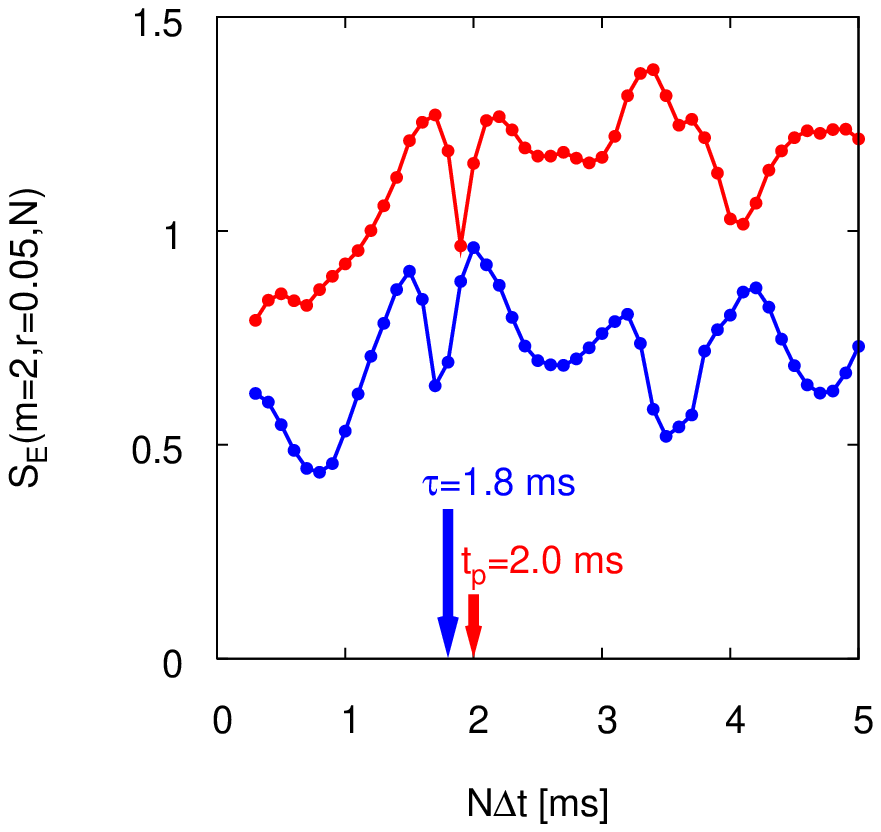}

  \parbox[b]{1mm}{(b)}\includegraphics[width=0.43\textwidth]{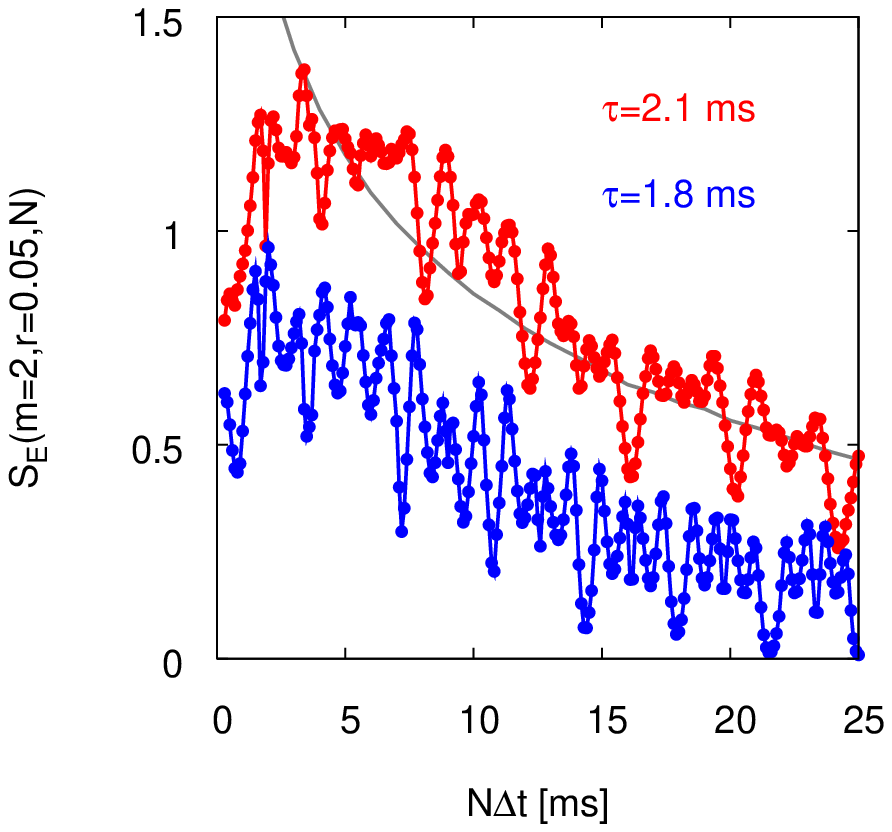}
  \caption{Multiscale entropy $S_E$ for fixed $m=2$ and $r=0.05$ depending on scale factor $N$. 
  	 The time series with delay time $\tau = 2.1$\,ms seems to be more complex
  	 than the time series with delay time $\tau = 1.8$\,ms since its entropy is larger.
  	 \textbf{a)} To gain a higher scale factor resolution we recorded also time series with smaller
  	 sampling time $\Delta t =0.1$\,ms.
  	 \textbf{b)} 
  	 The existence of a characteristic time scale which is in order of magnitude of the delay time 
  	 is indicated by the decay of entropy with increasing scale factor $N$. 
  	 The gray line represents GWN with $r=0.25$. 
  }
  \label{fig:MSE1d}
\end{figure}

To clear up the characteristics of the cutting process, we look at MSE depending on box size $r$
for the two chosen delay times, Fig.\,\ref{fig:MSE2d}.
For regular motion we expect the entropy to approach zero with decreasing $r$. 
This is observed for the time series with delay time $\tau = 1.8$\,ms.
For chaotic dynamics the entropy should stay finite, observed for $\tau = 2.1$\,ms.
For the sake of completeness, it should be mentioned that in the case of
stochastic dynamics the entropy would diverge with decreasing spatial resolution $r$, cp. Eq.\,(\ref{eq:MSE_GWN}).
In Fig.\,\ref{fig:MSE2d} and \ref{fig:MSE1d} we further analyze the scale factor dependence of MSE.
The entropic measure is always larger for the chaotic time series since it is the more complex one.
MSE for small scale factor, Fig.\,\ref{fig:MSE1d}a, indicates that there is no characteristic time scale, 
comparable to \mbox{$1\!/\!f$-noise} \cite{Costa2002}.
But for larger scale factors MSE is decaying comparable to Gaussian white noise, Fig.\,\ref{fig:MSE1d}b. 
Thus, even in the chaotic case, there exists a characteristic time scale which is close to the delay time.
The frequencies dominating $S(\omega)$, Fig.\,\ref{fig:PSD}, are also present in $S^{(N)}_E(m,r)$.
They belong to minima in Fig.\,\ref{fig:MSE1d}. We learn 
coarse-graining of the data over multiple of time scale $t_p$
belonging to structures in the cutting process dynamics
leads to less complex time series and contains less information.

\section{Conclusions and last remarks}
Concluding, the 0-1\,test differentiates between the two types of motion.
Depending on the chosen delay time for the investigated DDE, \mbox{Eqs.\,(\ref{eq1}) and (\ref{eq2})}, regular or chaotic motion is observed
and a transition from chaotic to regular motion is detected with increasing cutting speed.
The nature of solutions has been also confirmed by the corresponding
power spectral densities and multiscale entropies. 
The latter reveals more insights into the process dynamics 
but is of much higher computational cost
than the 0-1\,test and the spectral calculations.

The 0-1\,test appeared to be relatively simple 
and, consequently, useful for systems with delay and discontinuities.
A huge advantage of the test is its low computational effort and the possibility to compute 
it "on the fly" while the data is still growing.
One of the  useful aspects of
the 0-1 test is that the result can be plotted against the parameter $\tau$.

The presented method gives a quantitative criterion for chaos similar to the maximum Lyapunov exponent.

As demonstrated by Falconer {\em et al.} \cite{Falconer2007} and  Krese and Govekar \cite{Krese2012} the method can be used on experimental data as well.
Unfortunately in case of the cutting process experimental data are often characterized by a relatively high level of noise \cite{Litak2004}.
In the examined system, we waived the possibility of additive noise. 
It  was shown that the 0-1\,test could be applied on 
dynamical systems with additive noise and a good signal to noise ratio \cite{Gottwald2005}.

\section*{Acknowledgements} 
This work is partially supported by the European Union within the 
framework of the Integrated Regional Development Operational Program 
as project POIG.0101.02-00-015/08 and by the 7th Framework Programme FP7-REGPOT-2009-1,
under Grant Agreement No. 245479.

\end{document}